# Cation Discrimination in Organic Electrochemical Transistors by Dual Frequency Sensing

*Sébastien Pecqueur\*, David Guérin, Dominique Vuillaume and Fabien Alibart\**

S. Pecqueur, D. Guérin, D. Vuillaume, F. Alibart
Institut d'Électronique, Micro-électronique et de Nanotechnologie, CNRS, CS 60069, Av. Poincaré, 59652 Cedex, Villeneuve d'Ascq, France
E-mail: sebastien.pecqueur@iemn.univ-lille1.fr; fabien.alibart@iemn.univ-lille1.fr

In this work, we propose a strategy to sense quantitatively and specifically cations, out of a single organic electrochemical transistor (OECT) device exposed to an electrolyte. From the systematic study of six different chloride salts over 12 different concentrations, we demonstrate that the impedance of the OECT device is governed by either the channel dedoping at low frequency and the electrolyte gate capacitive coupling at high frequency. Specific cationic signatures, which originates from the different impact of the cations behavior on the poly(3,4-ethylenedioxythiophene):poly(styrenesulfonate) (PEDOT:PSS) polymer and their conductivity in water, allow their discrimination at the same molar concentrations. Dynamic analysis of the device impedance at different frequencies could allow the identification of specific ionic flows which could be of a great use in bioelectronics to further interpret complex mechanisms in biological media such as in the brain.

## 1. Introduction

Interfacing efficiently biology with electronic systems remains one of the biggest challenges of the coming decades. Potential applications ranges from the basic understanding of key biological processes in living systems to the possibility to replace, complement biological elements with electronic hardware. In particular, reaching this goal will require developing innovative approaches at the material and device level. In order to bridge biological medium, bioelectronics devices coupling electronic and ionic processes are used to implement complex features. To this end, OECT devices are attracting great attention as bioelectronics sensors [1–5]. In these structures, the biocompatible [6,7] semiconducting polymer ensures an efficient electrical transduction of an interfacing electrolyte ionic concentration into an electrical current. The bulk dedoping of the semiconductor upon gate field-effect being correlated to the ion concentration of the electrolyte [8], much effort is devoted for their application as conductimetric sensor. Furthermore, in order to discriminate different ions, strategies are employed to enhance the selectivity of the device [9–12]. Many of them are based on the synthesis and integration of materials with specific chemical functionalities, which increases the device structures complexity. If these approaches are efficient for single species detection, they remain limited for complex biological media analysis with various species and require co-integration of multiple ion-specific materials and devices. Here, we propose an innovative route that gather the high sensitivity of OECT and the possibility to use a single PEDOT:PSS-based device as a higher order chemical sensor, extracting multiple ion-specific outputs (i.e. on demand selectivity) [13]. Using the mixed ionic/electronic conduction of the device, we show that the ion-dependent impedance of OECT is limited by either an electronic or an ionic resistance, depending of the measurement frequency. Since the presence of the ions influences each resistance individually by different mechanisms, each output impedance can be decoupled the one from the other. The resulting two uncorrelated information allows cation discrimination at a given concentration, and therefore sense instantly, selectively and quantitatively cations, thanks to the two-dimensional feature of the OECT. We further apply this strategy to ion discrimination in a simple pulse-based sensing strategy that circumvents the utilization of complex impedance spectra recording.

## 2. Results and Discussion

### 2.1. Impedance Spectroscopy

*2.1.1. Typical Behavior*

Characterization of the frequency-dependent behavior of the PEDOT:PSS/electrolyte interfaces has already been extensively investigated [14–18] and shows either (i) the electrolyte-resistance dependency at high frequency [14–16] or (ii) the dedoped channel transconductance at low frequency [17,18]. These two different transient- and steady-state currents modeled by Bernards and al. [19] are governed by ion concentration limited mechanisms, which respectively increases the ionic conductivity, and decreases the OECT channel transconductance. In this study, we couple both of these uncorrelated and opposite trends to extrapolate further electrolyte information. First, we measured the impedance spectroscopy of four device geometries (Fig. 1a and 1b), in a $KCl_{(aq)}$ electrolyte, to determine the contribution of the channel and source/drain contacts. In each of the four cases (Fig. 1c and S1 in supplementary materials), the device behaves as a high-pass filter with an impedance of the following type Equation (1).

$$Z(f) = \frac{|Z|_{low} + j|Z|_{high}\frac{f}{f_0}}{1+j\frac{f}{f_0}} \quad (1)$$

with $|Z|_{low}$ the impedance modulus plateau at low frequency (typically f < 1 kHz), $|Z|_{high}$ the impedance modulus plateau at high frequency (typically f >100 kHz), and an impedance phase change at $f_0$ from $|Z|_{low}$ to $|Z|_{high}$ between 1 kHz and 100 kHz. Considering a rather low-complexity equivalent circuit (Fig. 1d and 1e), the system impedance can be written as Equation (2):

$$Z(f) = \frac{R_h + j\, R_M \cdot R_h \cdot C\, 2\pi f}{1 + j\, (R_M + R_h)\, C\, 2\pi f} \quad (2)$$

with $R_M$, the resistance of the electrolytic path, $R_h$ the resistance of the electronic path and the capacitance C. Using the definition of Equation (1), Equation (3):

$$Z_{low} = R_h \; ; \; Z_{high} = R_M \cdot \left(\frac{1}{1+R_M/R_h}\right) \; ; \; f_0 = \frac{1}{2\pi \, (R_M+R_h) \, C}$$

(3)

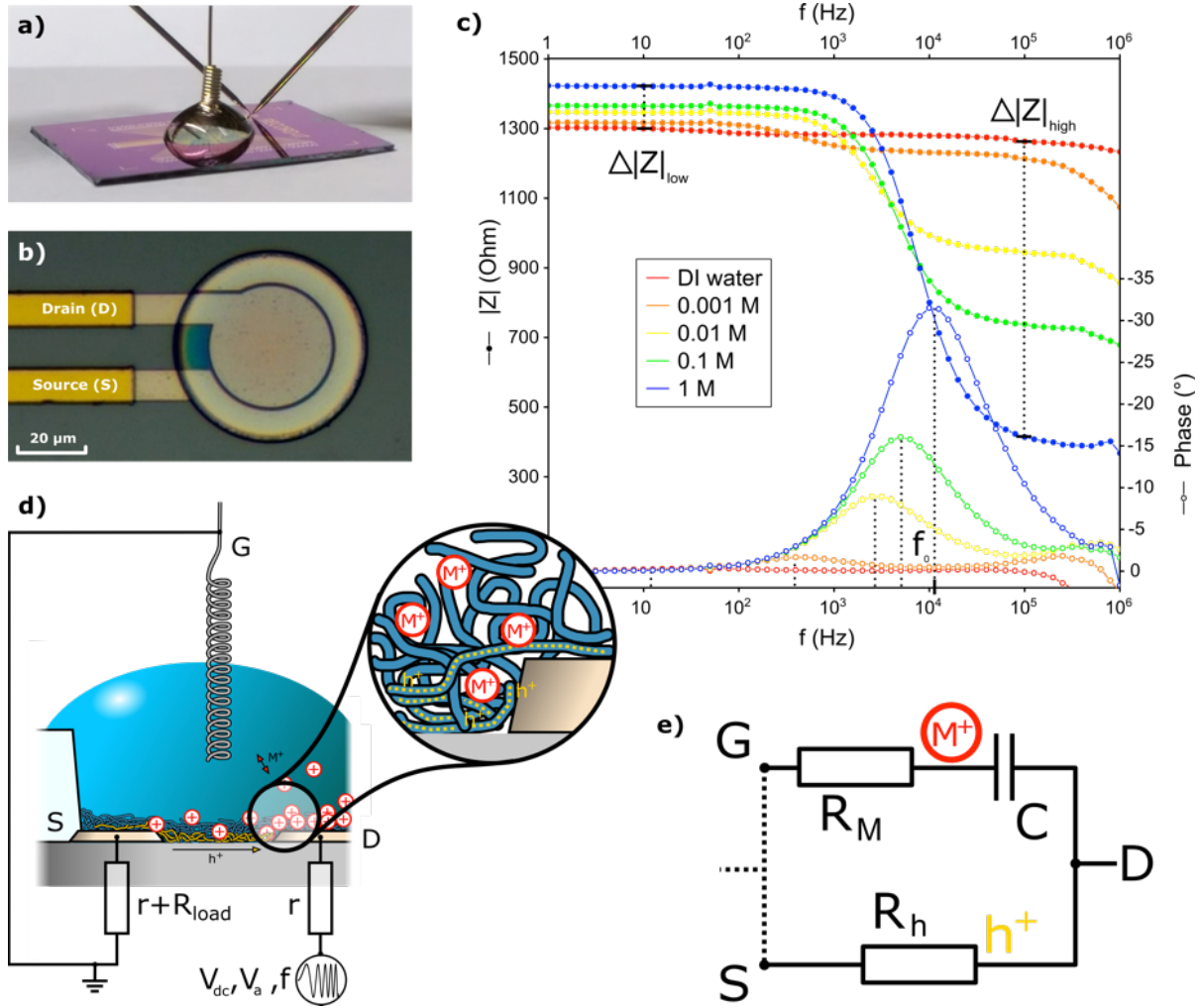

**Fig. 1.** (a) OECT chip under characterization. (b) Microscope picture of one of the concentric-electrode OECT (channel length L = 1 µm; drain diameter $R_{int}$ = 20 µm). (c) Typical trends for the impedance modulus and phase versus the electrolyte concentration, illustrated with the "long channel / large contact" OECT characteristics (L = 20 µm; $R_{int}$ = 50 µm; $V_{dc}$ = -50 mV). (d) Scheme of the device cross-section, electric setup for the impedance measurements, and charge-carrier mechanisms. $M^+$ (red circles) and $h^+$ (yellow dots) represent cations and holes, respectively. (e) Simplified equivalent circuit proposed to justify the observed impedance behavior.

The value of the low frequency plateau is exclusively correlated to the electronic resistor which depends on the PEDOT:PSS conductivity and the load resistor. The high frequency plateau is correlated to the electrolytic resistor (which depends on the electrolyte's conductivity) and to the electronic one only in the case of a not sufficiently conductive electrolyte (i.e. $R_M >> R_h$). The presence of the 1 kΩ load in the electronics resistor promotes the case $R_M << R_h$, and therefore, promotes the non-sensitivity of the high impedance plateau to the electronic resistor.

*2.1.2. Geometry Variation*

Considering Equation (2) that expresses the device impedance as a function of a capacitance, an electronic resistance and an electrolytic resistance, device geometry characteristics such as top area of the electrodes and inter-electrode distances must govern the sensitivity of the low and high frequency responses, and shall be optimized.

Experimentally for all device geometries tested, these plateaus show opposite trends with the concentration of electrolyte, such as the plateau value $|Z|_{low}$ increases of $\Delta|Z|_{low}$ with the increase of the concentration whereas $|Z|_{high}$ plateau value decreases of $\Delta|Z|_{high}$ with the increase of the concentration (data available in the supplementary data section in fig. S1). The opposite trends indicate different mechanisms, which are accessible individually at different frequencies, either promoting or lowering the electrical conduction.

At low frequency, the device impedance is dominated by the hole conduction through the PEDOT:PSS [17]. The concentration dependency of the measured impedance is representative of the increasing dedoped-state regime (with increasing cation concentration), as also accessible under DC measurements.

At higher frequencies, the charging of the interfaces predominates over the source-drain conduction and therefore the cations (which are charging the PEDOT:PSS/electrolyte interface), become the majority carriers with respect to the holes [14].

In this latter case, the device could be compared to a loosely short-circuited conductimetric cell for which the impedance decreases with the increase of the electrolyte concentration (we notice that the impedance can reach values below the load $R_{load}$, indicating that the main contribution of the current do not originate from the source electrode). The geometry-dependency on both modulations agrees on the proposed mechanisms for the impedance concentration sensitivity (Fig. 2). At a given concentrations, devices with the longest channel (L = 20 μm compared to 1 μm) show the largest $\Delta|Z|_{low}$ due to the channel dedoping, while devices with the largest drain electrode ($R_{int}$ = 50 μm compared to 20 μm) displays the largest $\Delta|Z|_{high}$ due to a larger capacitive coupling. Considering these observations, the following of the study has been focused on the "long-channel/large-drain" geometry (L = 20 μm; $R_{int}$ = 50 μm).

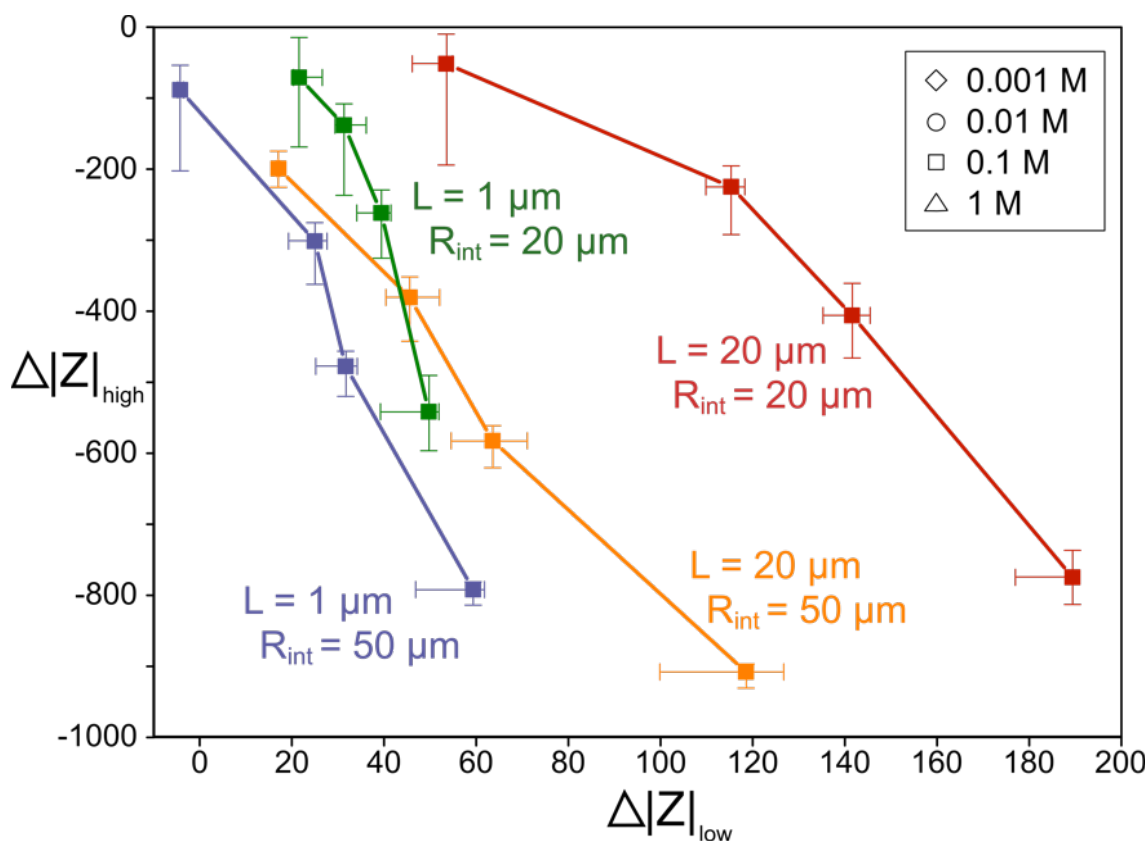

**Fig. 2.** Dependency of the OECT impedance-modulus change at low and high frequency with the device geometry (supplementary data available in fig. S1).

*2.1.3. Frequency-Dependent Ionic Specificity*

0-Order sensors delivering only single steady feature provides a single information, usually related to the quantity of species responsible for the sensing mechanism [13]. And in order to generate a higher order (providing a first degree of selectivity in addition of the sensitivity), arrays of different devices can be used. Using the OECT as a 1-Order sensor would provide both selectivity and sensitivity of the sensor without requiring populations of sensors. Therefore, OECT could provide a spatially confined qualitative and quantitative information on the ionic medium, if steady-state and transient features are ion specific. To this end, electrolyte samplings of $KCl_{(aq)}$, $LiCl_{(aq)}$, $CsCl_{(aq)}$, $CaCl_{2(aq)}$, $NaCl_{(aq)}$ and $RbCl_{(aq)}$ were prepared in ELISA plates from 1 M mother solutions to realize successive dilutions of $2^{-n}$ M solution (n from 0 to 11). We systematically analyzed the cation dependency of the OECT response from 1 to $5 \cdot 10^{-4}$ M and the impedance spectroscopy was performed under three different DC voltages: $V_{DC}$ = -50 mV, -350 mV and 550 mV (Fig. 3, S2 to S4 in supplementary materials).

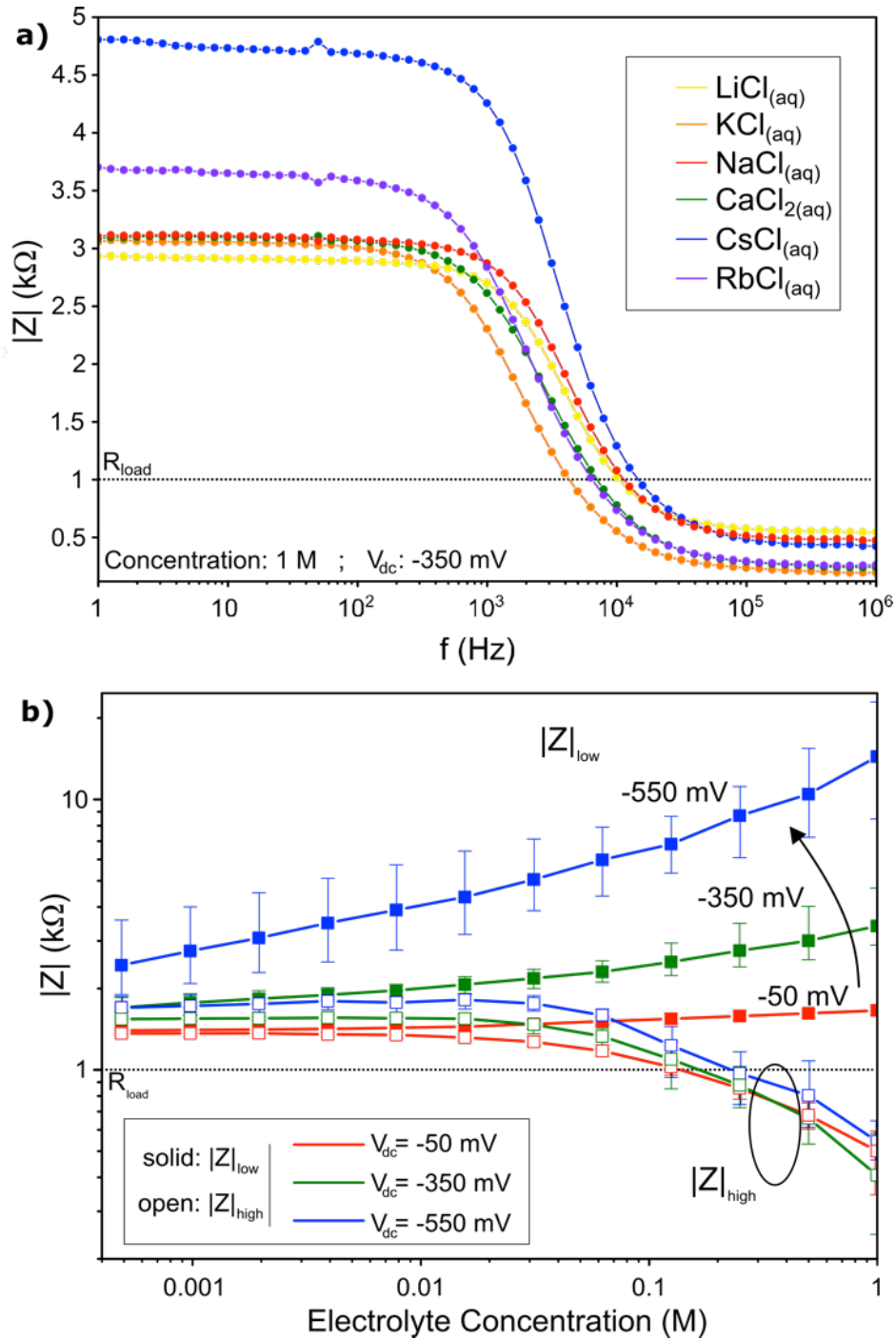

**Fig. 3.** (a) Impedance modulus data displaying a cation dependency at low and high frequency (concentration of 1 M and $V_{DC}$ = -350 mV). (b) Impedance modulus for the six different salts versus cation concentration, showing the DC voltage dependency at low frequency.

For each electrolyte, the impedance behavior with the concentration and over the frequency is very comparable to the one previously displayed for $KCl_{(aq)}$: the high impedance plateau at low frequency $|Z|_{low}$ increases with the concentration while the low impedance plateau at high frequency $|Z|_{high}$ decreases with the increase of the electrolyte concentration (Fig. 3a). It shows that the nature of the metal chloride salt used in this set of experiments is compatible with the operation of the OECT: all the six cations support the dedoping of the PEDOT:PSS and contribute to the capacitive coupling. Nevertheless, the values of the two plateaus depend on the electrolyte nature at the same concentration, denoting a cationic dependency of the impedances ruled by both mechanisms (Fig. 3a).

In all cases, the impedance is DC-voltage dependent at low frequency while not at high frequency (Fig. 3b). For example, in the case of the $CsCl_{(aq)}$ electrolyte, $|Z|_{low}$ increases by an order of magnitude: from 1.7 k$\Omega$ (at $4.9 \cdot 10^{-4}$ M) to 23 k$\Omega$ (at 1 M) at 550 mV. This observation confirms that the impedance at low frequency is dominated by the doping state of the PEDOT:PSS channel, modulated by the difference of potential between the gate and the PEDOT:PSS [19]. At high frequency, the value $|Z|_{high}$ for each series of salt/concentration remains unchanged with $V_{DC}$ by an averaged standard deviation of 14% of the mean value of $|Z|_{high}(V_{DC})$. This shows that the limiting mechanism governing the device impedance at high frequency does not depend on the gate potential. Therefore, it confirms that the ion dependency observed both at high and low frequency are yielded by two different mechanisms. Also, considering the low-frequency impedance is dominated by the cation dedoping strength while the high-frequency impedance is dominated the electrolytic conductivity, both impedance values potentially deliver two uncorrelated ion-related outputs. To verify this assumption, low- and high-frequency $\Delta|Z|$ values for the different electrolytes have been plotted versus concentrations (Fig. 4).

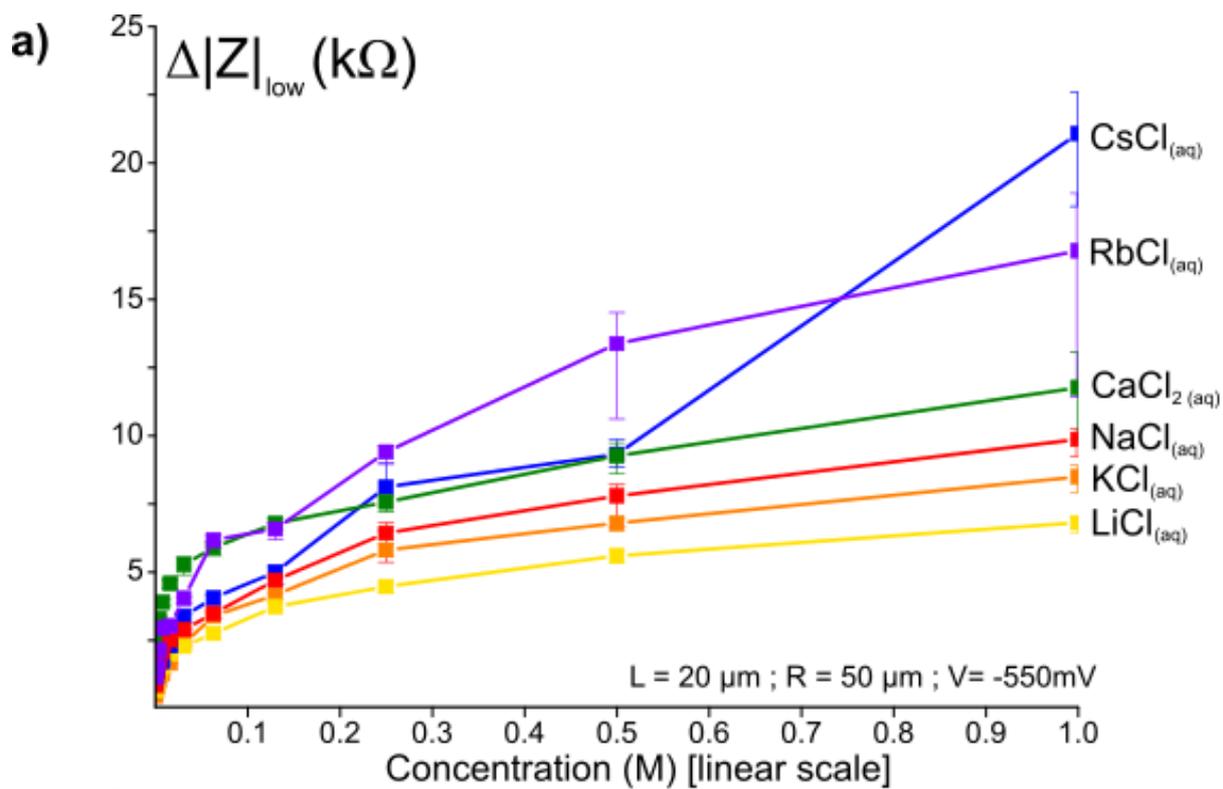
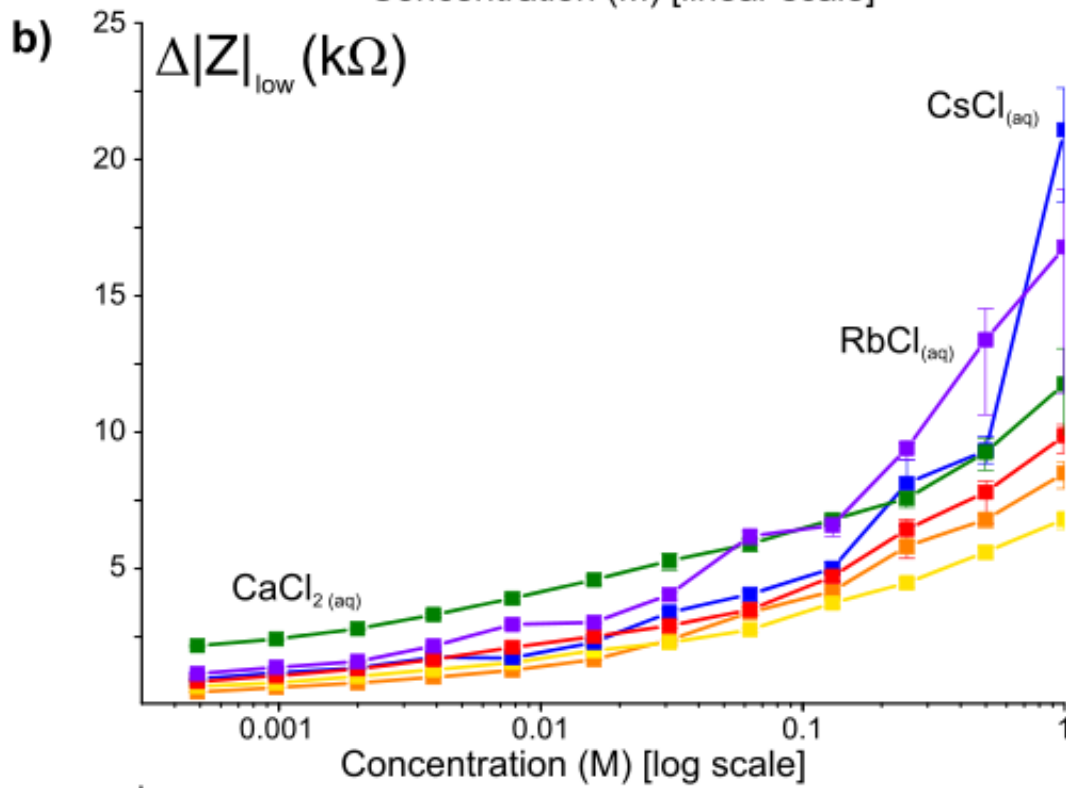

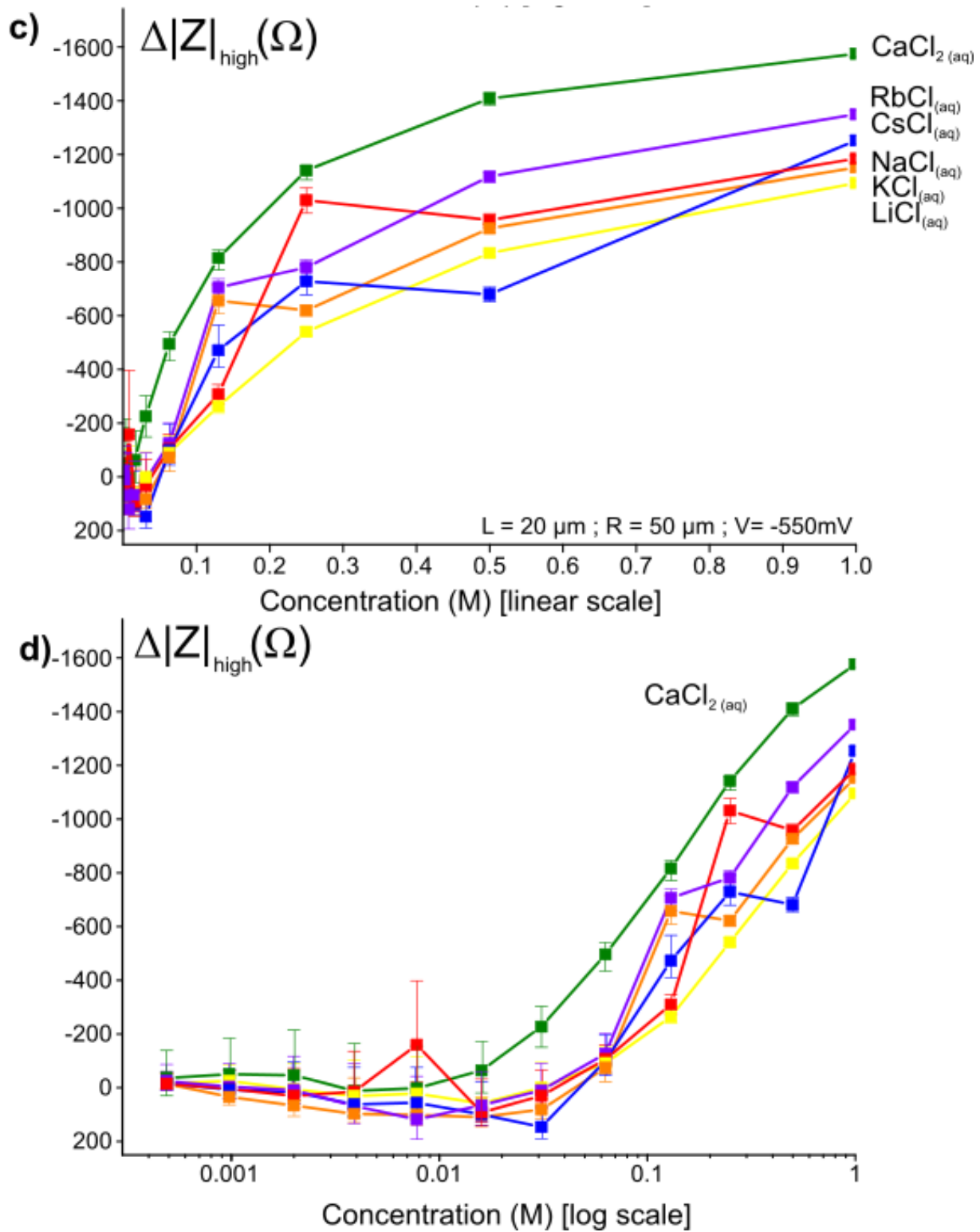

**Fig. 4.** Concentration dependency for the low-frequency impedance modulus variations (a and b) and high-frequency impedance modulus variations (c and d) of six different cations chloride aqueous electrolytes under $V_{DC}$ = -550 mV (data at -350 mV and -50 mV available in the supplementary data section, fig. S5 and S6). Abscissa in linear scale (a and c) and logarithmic scale (b and d) for the ion concentration.

On the low-frequency measurements at $V_{DC}$ = -550 mV (Fig. 4a and b), we noticed that the ion dependencies of the devices are different at low and high concentrations. This feature is also observed for the data at $V_{DC}$ = -350 mV, (Fig. S5 in the supplementary materials). At concentrations higher than 0.1 M (Fig. 4a), the increase of impedance modulus is comparatively higher for electrolytes with high atomic-number/size cations (with the lowest impedance values for $LiCl_{(aq)}$ and the highest ones for $CsCl_{(aq)}$ and $RbCl_{(aq)}$). At concentrations lower than 0.1 M (Fig. 4b), the impedance modulus variation is higher for the divalent cation, compared to the monovalent ones (observed only in the case of $V_{DC}$ = -550 mV).

At high frequencies, the data (Fig. 4c and d) are less reproducible from one concentration to the other (probably due to the lower variations $\Delta|Z|_{high}$ compared to $\Delta|Z|_{low}$). Nevertheless, a clear trend is observed at $V_{DC}$ = -550 mV (barely observed for $V_{DC}$ = -350 mV and not observed in $V_{DC}$ = -50 mV) along the whole range of concentrations for the $CaCl_{2(aq)}$ electrolyte to conduct more than the alkali ones. This observation agrees with an ionic-conductivity driven mechanism (or diffusion control as proposed by Coppedè et al. [11], since both coefficients are proportional) at high frequency, limited by ion mobility but also by the valence of the involved cations.

*2.1.4. Cation Discrimination*

Both static and dynamic responses of the OECT have different ion specificity, as the ionic series for impedance modulation variation differs at low and high frequency. To verify the non-coplanarity of the series of data-points and their separability over a two-dimensional space ($\Delta|Z|_{low};\Delta|Z|_{high}$), fig. 5a shows a three-dimensional plot of the data of the fig. 4a and 4c. We observed that the group of six cation series is not coplanar in any two-dimensional space: We noticed in fig. 5a that the data set for the divalent electrolyte is further frontward (toward lower $\Delta|Z|_{high}$) than the five series of other electrolytes (consistent with the data for $V_{DC}$ = -350 mV but not with $V_{DC}$ = -50mV, in the supplementary data section S7 to S9). From the lack of coplanarity, each series of electrolyte can be spatially discriminated at a given concentration, as illustrated in fig. 5b. Centered on their concentration-dependent average (cross-point at the center of the graph), the stacking of planes for 1 M, 0.5 M, 0.25 M and 0.125 M, shows localization of the individual salt in an ion-specific of the ($\Delta|Z|_{low}$;$\Delta|Z|_{high}$) plane. With data performed at the different DC biases at 350 mV and 50 mV (Fig. S7-S9 as supplementary material), larger DC biases promote better cation separations (for which $\Delta|Z|_{low}$ components are larger). Therefore, it is possible to extrapolate the cation's nature by the evaluation of both $\Delta|Z|_{low}$ and $\Delta|Z|_{high}$, knowing the relative position of the cations in this two-dimensional space map.

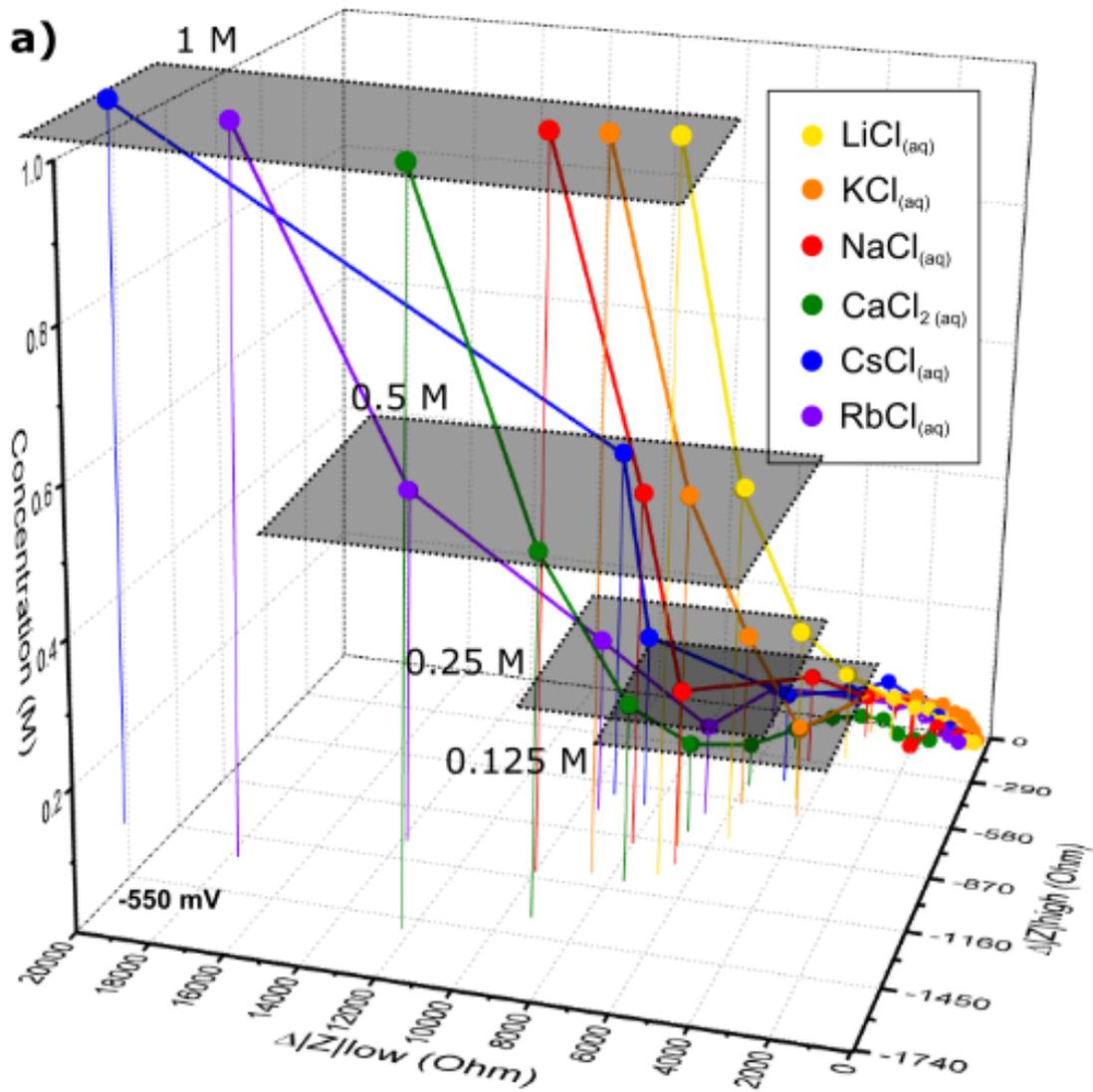

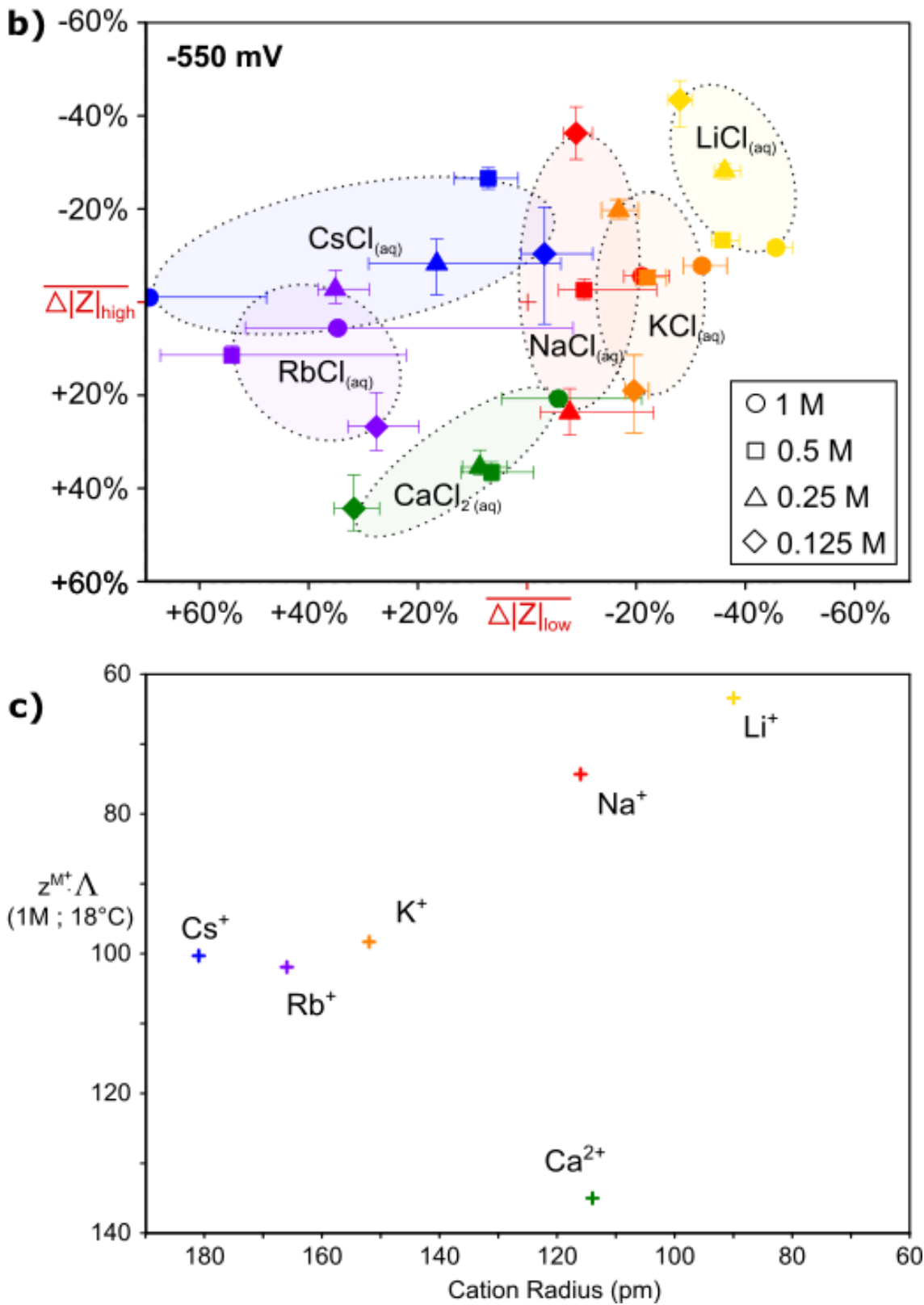

**Fig. 5.** (a) Three dimensional plot of the low- and high-frequency impedance modulus variations with the ionic concentration under $V_{DC}$ = -550 mV (data at -350 mV and -50 mV available in the supplementary data section, fig. S7 to S9). (b) Stacking of 1 M, 0.5 M, 0.25 M and 0.125 M planes, centered on their respective average point (cross point). (c) Representation of the cation with their cation radius (from Ref. [20a]) and the ionic conductivity in water (from Ref. [20b]), showing some analogy with the 1 M plane above.

An analogy is proposed in fig. 5c for the relative position of the impedance modulus modulation points ($\Delta|Z|_{low}$; $\Delta|Z|_{high}$) with the coordinates (cation radius ; electrolyte conductivity). The correlation between $\Delta|Z|_{high}$ and the electrolyte's conductivity derives, (i) from the diffusion limitation in transient regime already assessed by Coppedè et al. [11] and (ii) the linear relationship between the diffusion coefficient and the limiting ionic conductivity, by the Nernst-Einstein equation. For the proposed correlation between $\Delta|Z|_{low}$ and the size of the cation, the overall trend suggests a dependency on the dedoping strength of the cation with a cation-size dependent parameter ($LiCl_{(aq)}$ and $NaCl_{(aq)}$ being systematically less conductive at low frequency than $CsCl_{(aq)}$ and $RbCl_{(aq)}$). But no strict linear dependency between the cation size and the modulation $\Delta|Z|_{low}$ is assessed (particularly for $KCl_{(aq)}$). Anyway, we demonstrated experimentally that there is genuine selectivity (or specificity) in the standard PEDOT:PSS to be dedoped by the "bulkiest" cations.

To demonstrate the applicability of the discrimination by impedance spectroscopy, we focused the rest of this study on three cations that have very distinctive behaviors at low and high frequency: $NaCl_{(aq)}$ as a small monovalent cation, $CsCl_{(aq)}$ as a large monovalent cation and $CaCl_{2(aq)}$ as a divalent cation.

### 2.2. Application for a Dynamic Cation Recognition

Impedance spectroscopy provides an instant information of long-lasting time-dependent current responses over discrete frequencies. To transfer the impedance properties on a practical platform, the following application was conceived for dynamic cation recognition, providing an analog current modulated output by applying digital voltage inputs. A constant drain voltage ($V_D$) and a square-pulsed gate voltage ($V_G$ - duty cycle of 50%) were applied on the same OECT while we recorded the drain current (Fig. 6a). The shape of voltage bias was chosen to emulate the one that could be digitally addressed on conductimetric biosensors. The voltage magnitude at the gate (+550 mV); and the drain (-50 mV) electrodes were set in order to be in a comparable case as in the impedance spectroscopy measurement at $V_{DC}$ = 550 mV. Two square wave frequencies were used: 200 Hz (f< $f_0$ for 0.1 M solutions, quasi-steady) and 20 kHz (f> $f_0$ for 0.1 M solutions, in the transient). The device was sequentially exposed, in chronological order, to: (i) 0.1-M concentrated $NaCl_{(aq)}$, (ii) $CaCl_{2(aq)}$ and (iii) $CsCl_{(aq)}$; three times each salt from different electrolyte preparations and rinsed with deionised water between each step.

Taking the equivalent circuit previously considered for the impedance spectroscopy study (Fig. 1e), applying individual potential at the gate and drain (Fig. 6a) instead of bridging both electrodes, the time dependent drain current for squared pulse inputs is Equation (4):

$$ \qquad (4) $$

At low frequency (f < $f_0$), the current approaches the steady-state Equation (5):

$$ \qquad (5) $$

That means that $I_{low}$ is exclusively function of the electronic resistor, i.e. the doping state of the PEDOT:PSS ($I^+_{low}$ when doped and $I^-_{low}$ when dedoped).

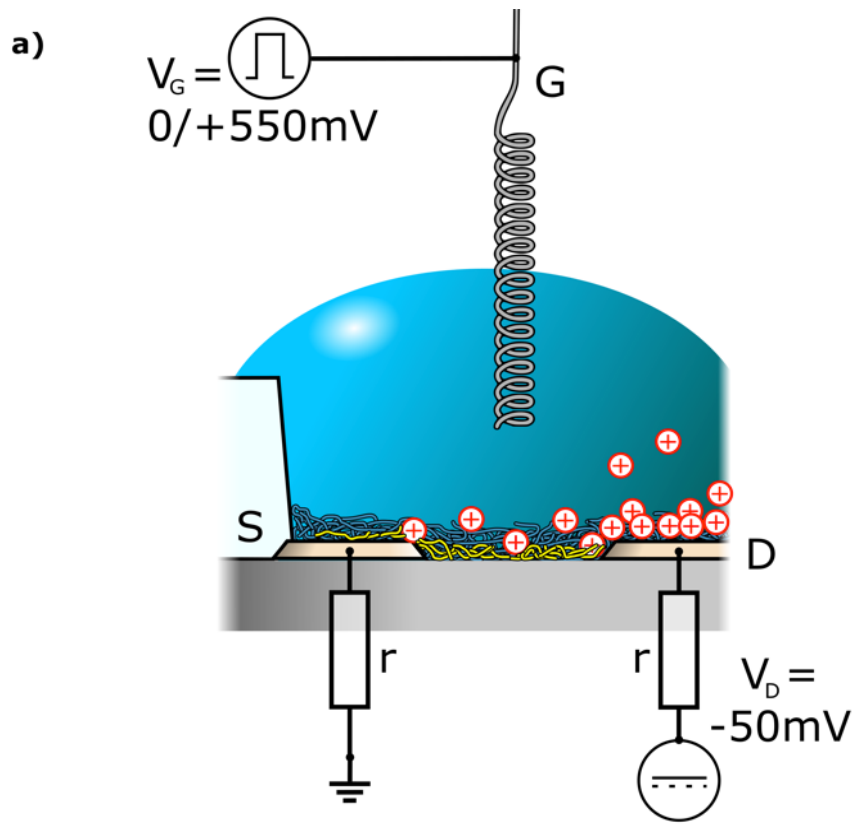

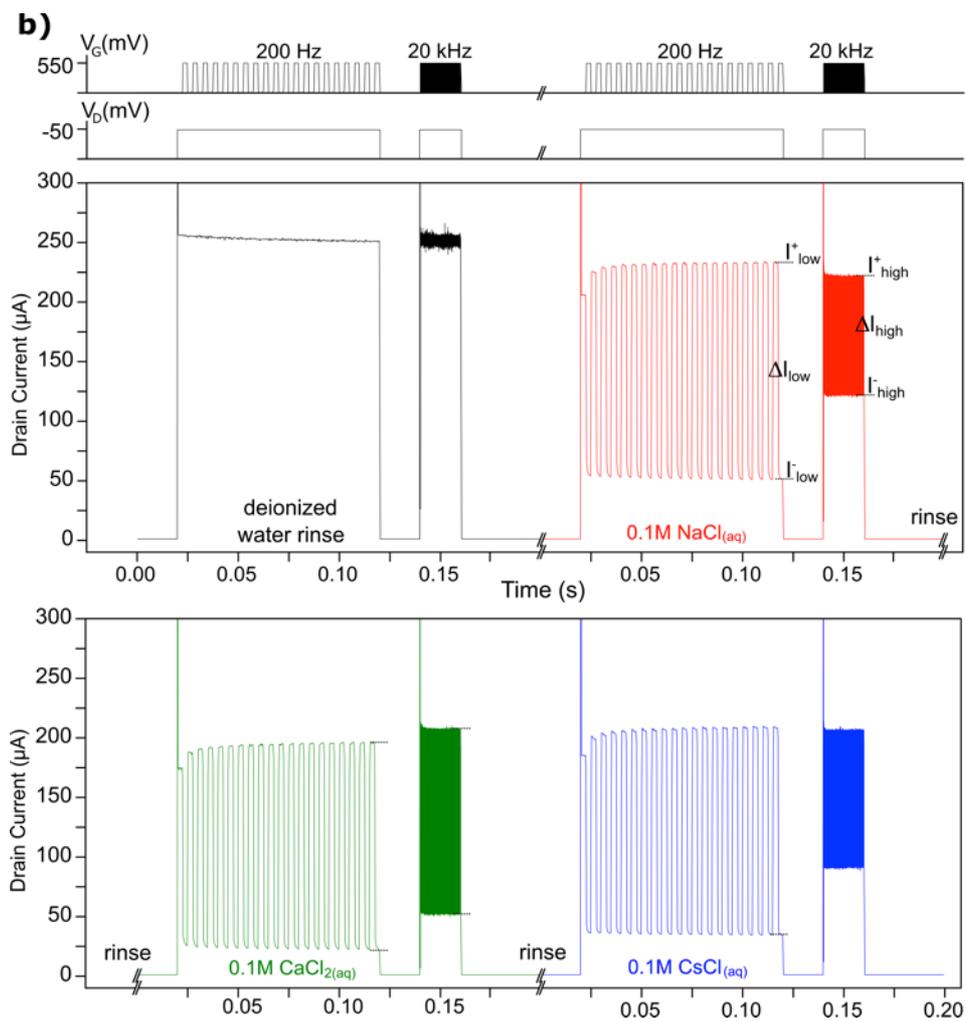

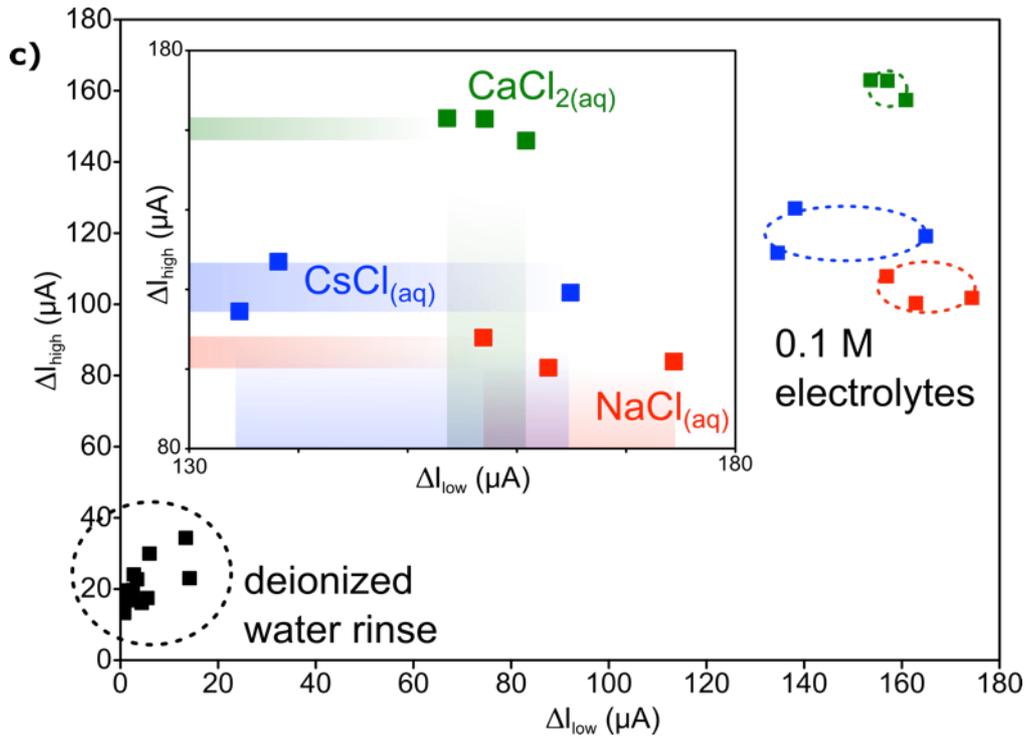

**Fig. 6.** (a) Electrical setup for squared pulses stimulation. (b) Output drain current for the OECT exposed at 0.1 M concentrated $NaCl_{(aq)}$, $CaCl_{2(aq)}$ and $CsCl_{(aq)}$ electrolytes. (c) Current modulation at low and high frequencies showing both a spatial discrimination between the presence and the absence of salt and also depending on the salt.

At high frequency (f > f₀), the exponential decay of the current approximates the tangential at the origin Equation (6):

$$ \quad (6)$$

As for the impedance spectroscopy study, the high frequency response depends partially on the electronic resistor $R_h$. Instead of adding a resistive load, we promoted the electrolytic contribution in the output current by applying a larger polarization ($V_G$-$V_D$) through the electrolyte rather than through the channel ($V_D$), leading to Equation (7):

$$ \quad (7)$$

Under these conditions, $I_{high}$ results entirely from the capacitive coupling through the electrolyte, limited by the resistance $R_M$.

Experimentally (Fig. 6b), the current is non-sensitive to the gate voltage pulse upon deionized water exposure, and equals 250 µA (a -20% drift overtime is observed during the multiple rinsing) at low and high frequency. This agrees with the trend displayed in fig. 1c for the impedance in deionized water, which behaves like a resistor. When exposed to the salts, different current modulations ΔI occurs between the maximum current $I^+$ (at $V_G$ = 0 V) and the minimum $I^-$ (at $V_G$ = +550 mV) at both low and high frequencies. Considering the expressions of equation 5 and equation 7, the different current modulations ΔI at high and low frequencies shall provide access to the cation discrimination similarly to the impedance spectroscopy measurement presented below.

This is confirmed in fig. 6c, where the current modulation coordinates for the different measurements are plotted in a ($\Delta I_{low}$;$\Delta I_{high}$) space. One can notice that the current modulations performed after rinsing the device in deionized water belong to a group very localized at the origin of the graph. For the 0.1-M concentrated solutions, three clusters of data points are localized on the upper right of the graph. We note that the three points for $CaCl_{2(aq)}$ are very localized in this graph and can clearly be separated from the rest of the monovalent cation salts. We note also a separation between $NaCl_{(aq)}$ and $CsCl_{(aq)}$ on the abscissa (one point of the cesium salt being a bit off the others). When looking at the ordinates of the three clusters, one can see that their distributions are very narrow and follow the series predicted by the impedance spectroscopy. On the abscissa of the three clusters, the distributions between the different groups is larger and overlap the one with the other. The reasons that could justify are (i) the observed -20% current drift (although we observed a special care for rinsing the device between each step), and (ii) the $\Delta I_{low}$ data points being dispersed only over 40 µA (63 µA for $\Delta I_{high}$). The separation of the 3 series of data points is anyway satisfying for the perspective of recognizing cations by valency or size over larger statistics of measurements.

An improvement strategy for diminishing the current drift, potentially due to cation remanence (from slow diffusion), could be by adding an extra pulse sequence after each step in order to electrostatically repeal the cation out of the PEDOT:PSS. Also, increasing the channel length L should promote the increase of $\Delta I_{low}$ and therefore the separability of the data points.

## 3. Conclusion

We confirmed the operability of the OECTs on a broader range of cations other than the extensively studied $K^+$ and $Na^+$. First, we showed that the impedances of the OECT device depend on both (i) the concentration and (ii) the nature of these electrolytic cations. Secondly, we evidenced that the device is more sensitive to large size/mass cations at low frequency, while it is more sensitive to the valency and the ionic conductivity at high frequency. We demonstrated that it is possible to selectively detect the cations present in the electrolyte by using a single device without specific chemical functionalization. Finally, we proposed a practical pulse protocol that confirms the possibility to recognize cations at the 100-ms scale, which could be applied to digitally address OECT biosensors for dynamic cation recognition. These results promote the discrimination of ion signals in biological fluids and could help applying OECT biosensors to unravel dynamically and more systematically biological mechanisms.

### 4. Experimental Procedures

*Device Fabrication:* The fabrication of the concentric electrode OECTs was based on methods reported elsewhere [12,21]: We first patterned on a Si/SiO$_2$ substrate the 70 nm Pt source and drain electrodes, and then the 300 nm Au electrodes (with a 10 nm Ti adhesion layer) by electron beam lithography and lift-off with a PMMA/MAA resist. Pt was chosen as a metal for the source and drain electrodes for its high work-function for an optimal charge-carrier injection, while gold was used to reduce the resistance of the transmission lines on the substrate (r = 46±5 Ω). After UV-O$_3$ cleaning, the substrate was functionalized with 3-acryloxypropyl trimethoxysilane in o-xylene (with 1 % acetic acid) before processing a 2-μm-thick CVD-processed parylene C passivation layer. To realize the semiconductor shadow mask, we spin-coated the Micro-90 surfactant over the substrate before processing another 2 μm parylene C passivation layer and dried-etched round apertures over the source, drain and channel areas with an oxygen plasma (30 sccm, 100 mTorr, 150 W). The PEDOT:PSS-based formulation was spin-coated over the substrate, and soft baked (115°C, 15 min) before peeling the parylene C mask (the electrodes were pretreated with a mercaptoethanol solution to promote the adherence of the PEDOT:PSS patches on the Pt electrodes). The thickness of the PEDOT:PSS was evaluated to be of 16±5 nm by profilometry. A hard baking (140°C, 60 min) and soaking overnight in deionized water was performed before use.

*Electrical Characterization:* The electrochemical impedance spectroscopy was measured on a Solartron Analytical ModuLab 2100A, applying a negative DC voltage bias V$_{DC}$ at the drain and bridging the source to the grounded gate. A shielded resistor load ($R_{load}$ = 1 kΩ) was placed in the source-gate bridge to promote a gate-drain electric field larger than the source-drain hole-drift field upon analysis. A small AC voltage ($V_a$ = 50 mV) at frequencies f from 1 Hz to 1 MHz is also applied for the impedance measurement. The squared voltage inputs were applied via two Agilent B1530A waveform generator units. All measurements were performed using an Ag coiled wire as a gate, dipped into a 150 μL electrolyte drop over the OECTs on the substrate. Although no noticeable influence on the data has been observed, the position of the coil and the volume of electrolyte were reproductively controlled for the sake of performances reproducibility.

**Acknowledgements**

The authors wish to thank the European Commission (H2020-FETOPEN-2014-2015-RIA, project RECORD-IT, #664786) and the French National Nanofabrication Network RENATECH for financial supports. We thank also the IEMN cleanroom staff of their advices and support.

**Appendix A: Supplementary data**

Supplementary data related to this article can be found at….

**Keywords: OECT, PEDOT:PSS, impedance spectroscopy, cation sensor.**

**Highlights:**
- Selective cation sensing from PEDOT:PSS-based Organic Electrochemical Transistors.
- Use of the OECT as a multi-parametric sensor and its frequency dependent impedance.
- Application for the discrimination of cations by square-wave frequency modulation.

**Graphical Abstract**

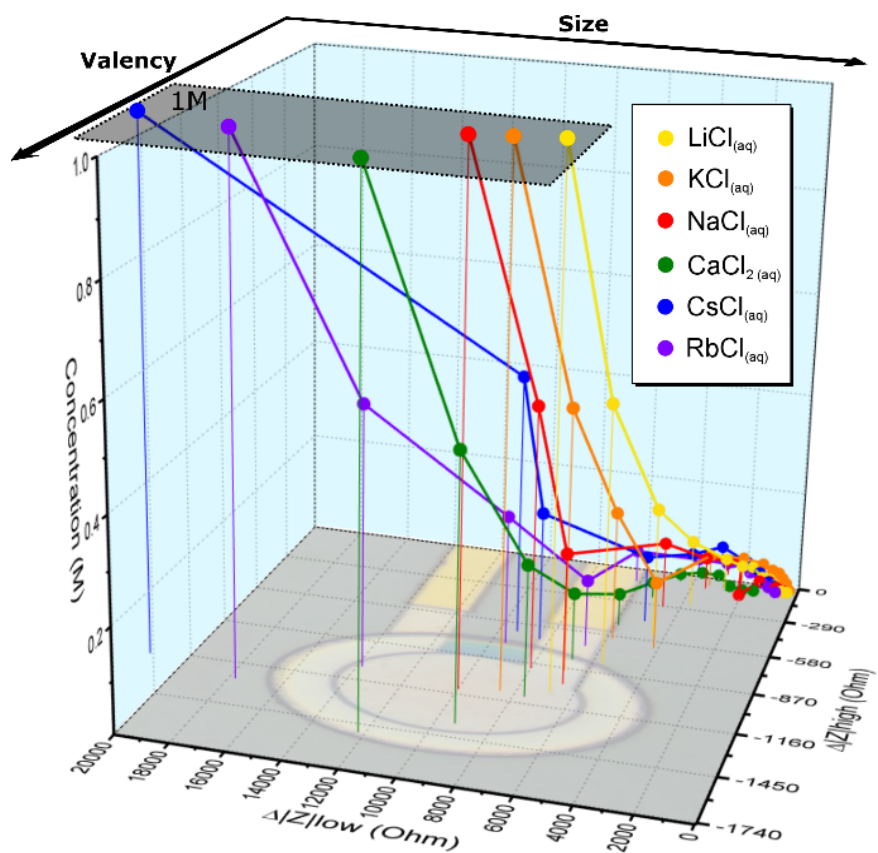

Supplementary data

**Cation Discrimination in Organic Electrochemical Transistors by Dual Frequency Sensing**

*Sébastien Pecqueur\*, David Guérin, Dominique Vuillaume and Fabien Alibart\**

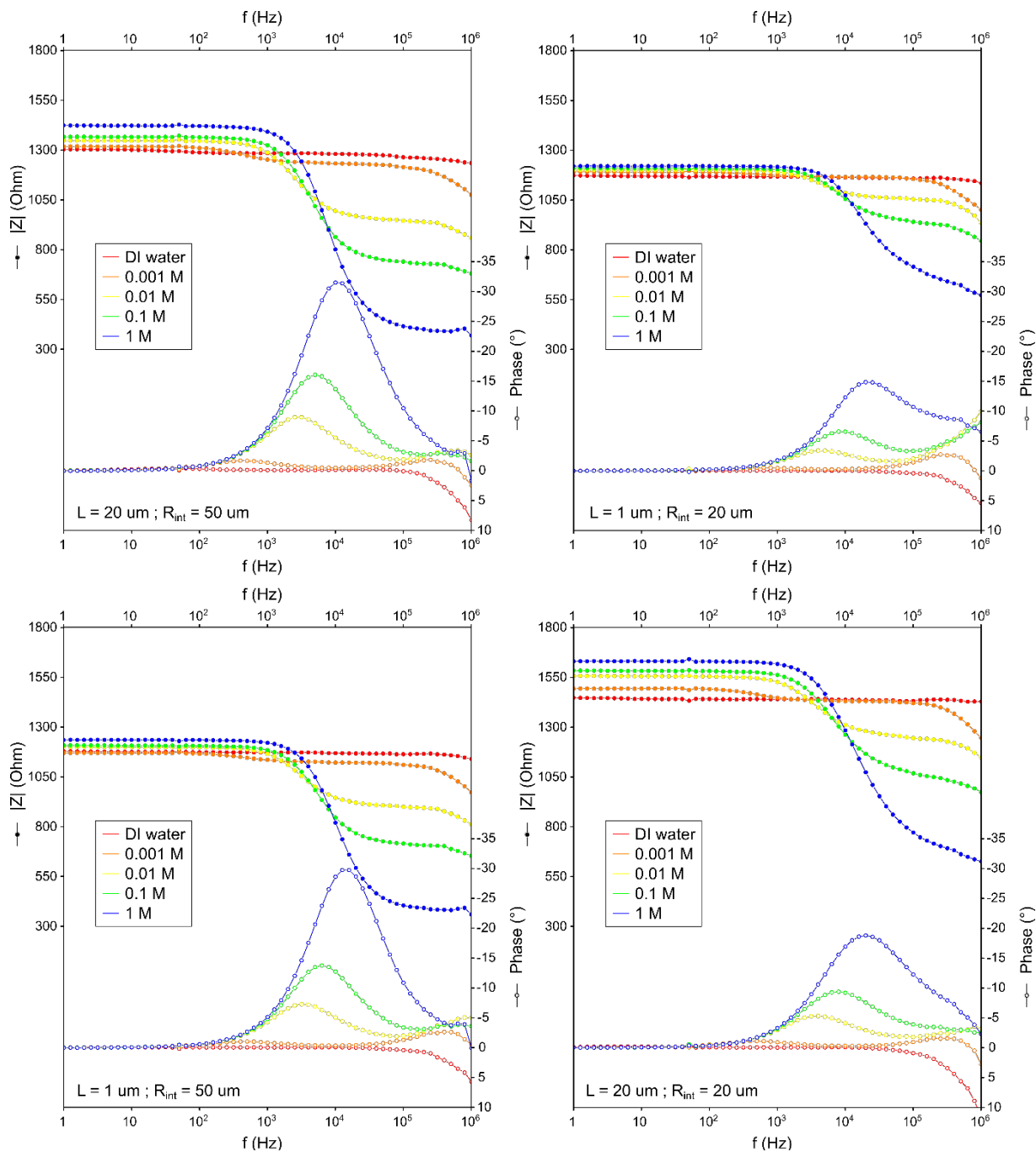

**Fig. S1**. Impedance modulus and phase measurement for four different device geometries, and at five different $KCl_{(aq)}$ molar concentrations

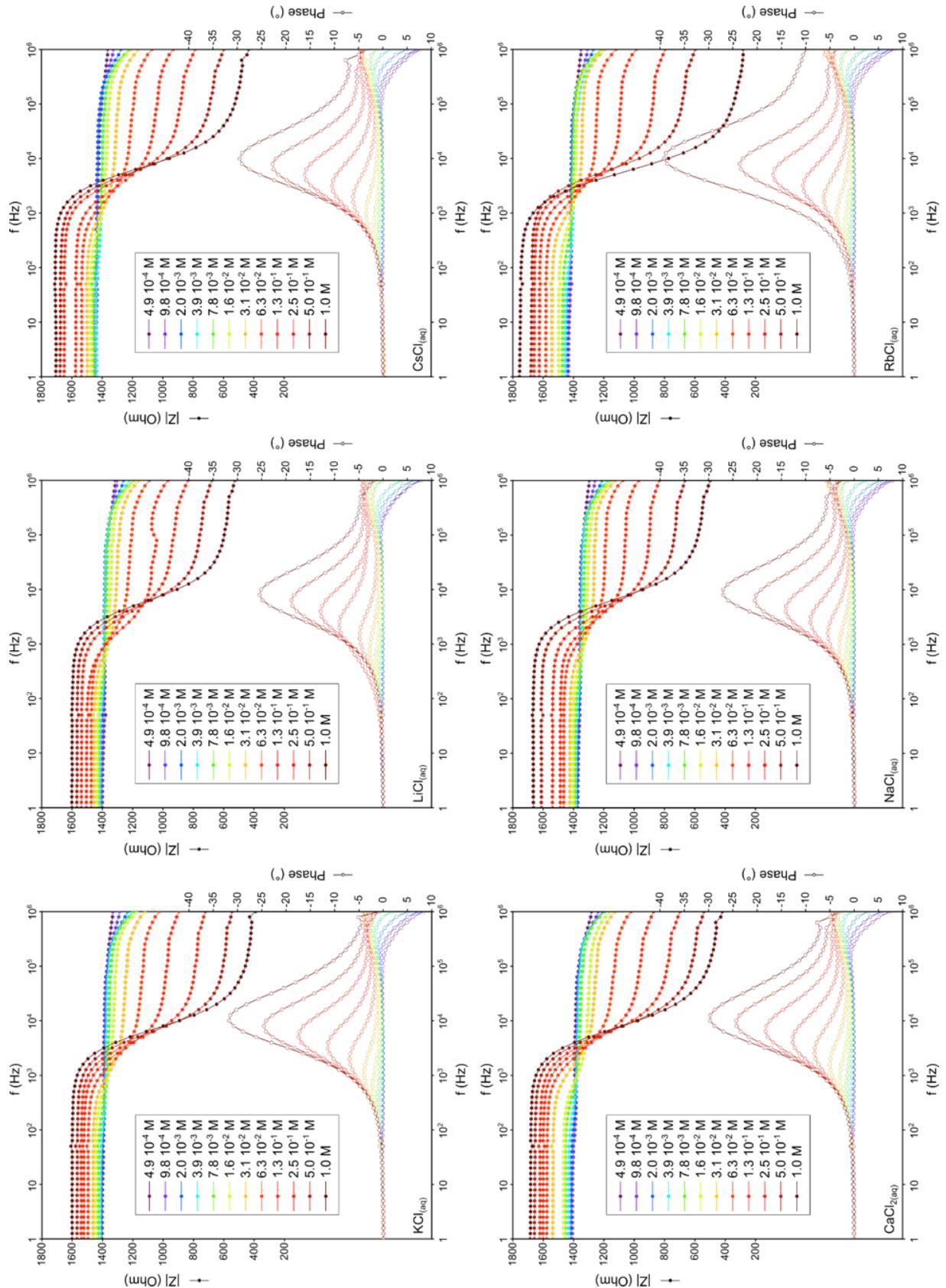

**Fig. S2**. OECT device ($R_{int}$ = 50 μm, L = 20 μm) impedance modulus and phase measurement, with six different cation chloride aqueous electrolytes at the concentration $1/2^n$ M (n from 0 to 11) at a stress of DC level = -50 mV and amplitude 50 mV at the drain, 1 kΩ-loaded source bridged to grounded gate.

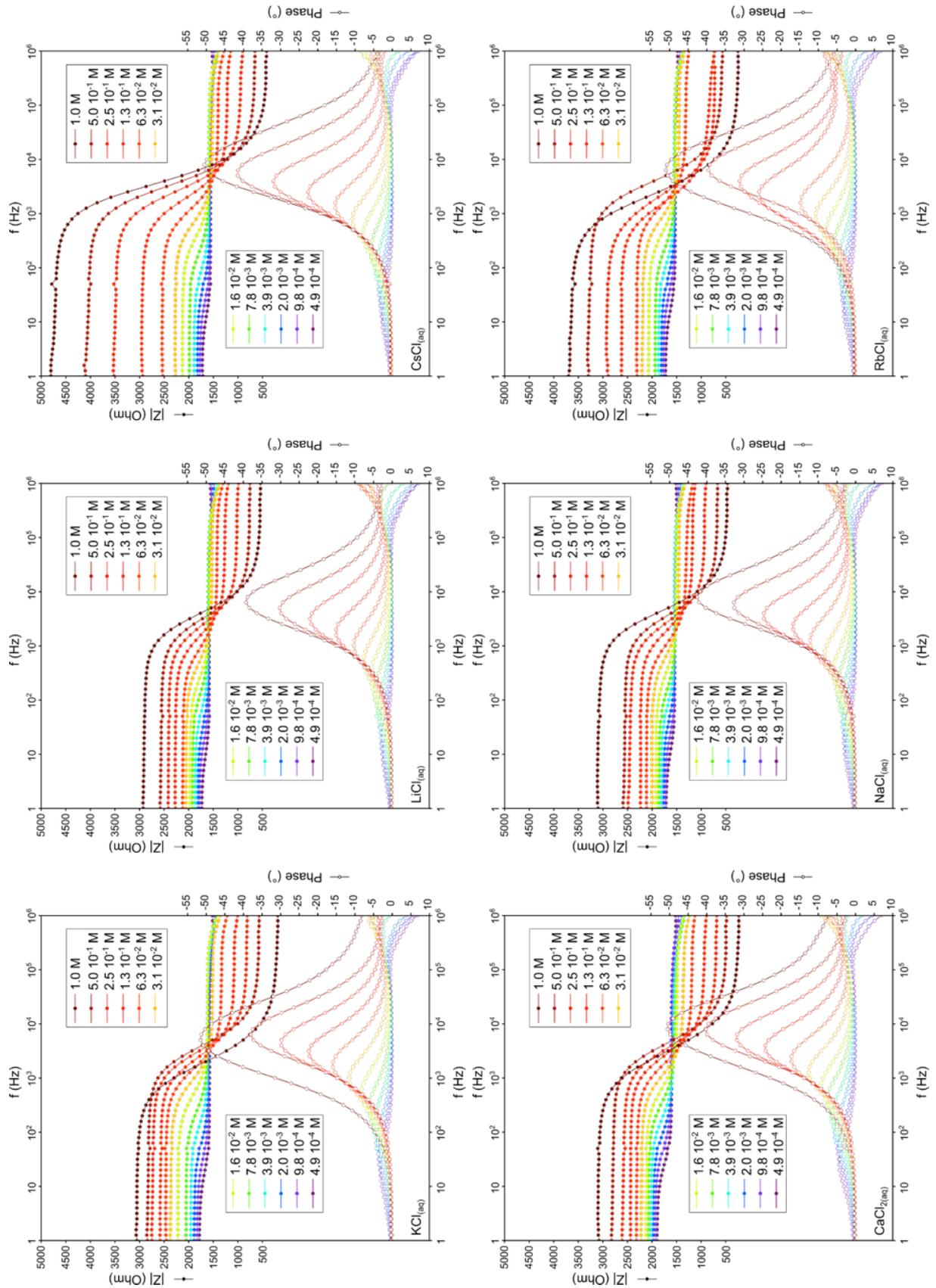

**Fig. S3**. OECT device ($R_{int}$ = 50 μm, L = 20 μm) impedance modulus and phase measurement, with six different cation chloride aqueous electrolytes at the concentration $1/2^n$ M (n from 0 to 11) at a stress of DC level = -350 mV and amplitude 50 mV at the drain, 1 kΩ-loaded source bridged to grounded gate.

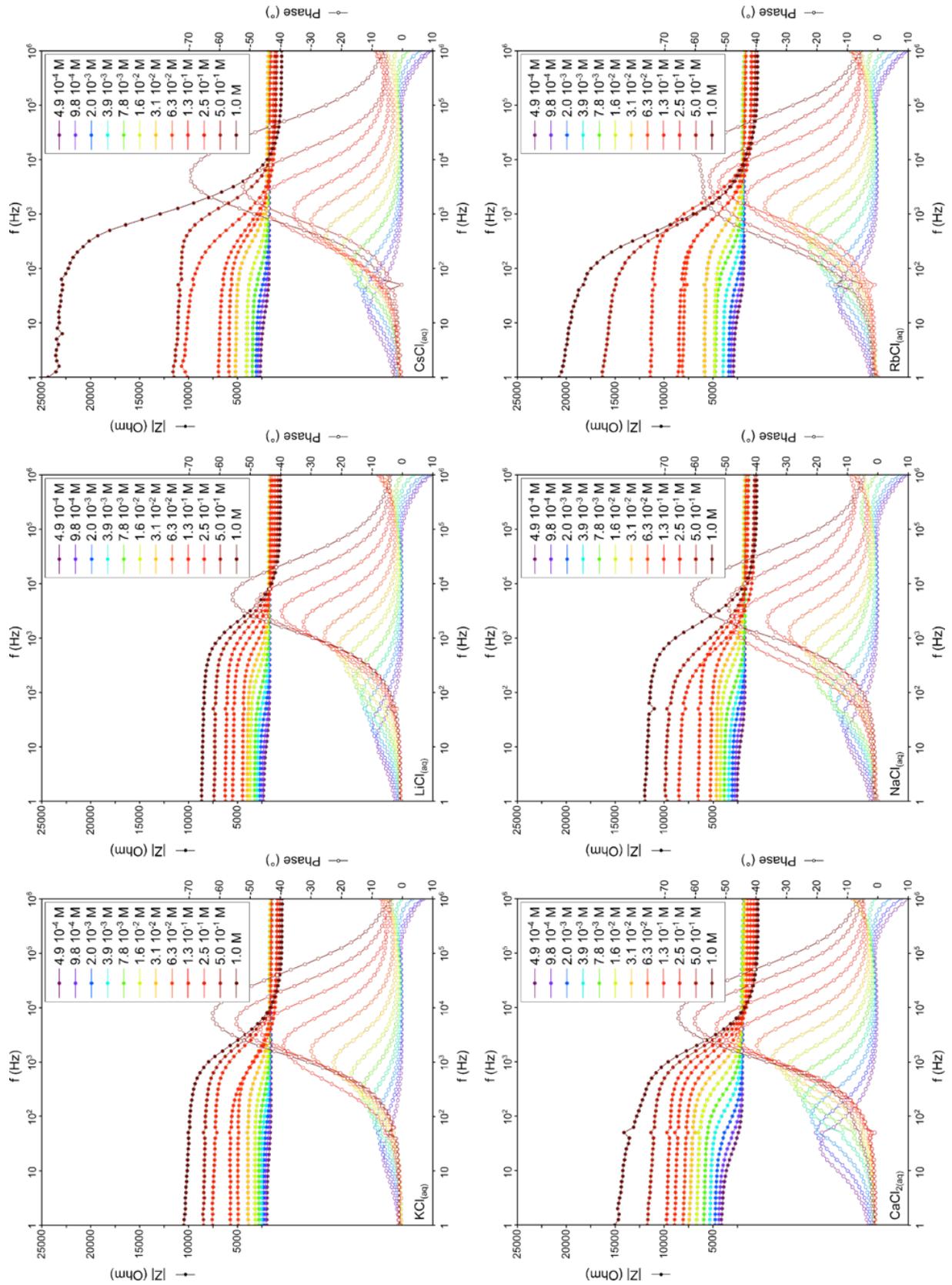

**Fig. S4**. OECT device ($R_{int}$ = 50 μm, L = 20 μm) impedance modulus and phase measurement, with six different cation chloride aqueous electrolytes at the concentration $1/2^n$ M (n from 0 to 11) at a stress of DC level = -550 mV and amplitude 50 mV at the drain, 1 kΩ-loaded source bridged to grounded gate.

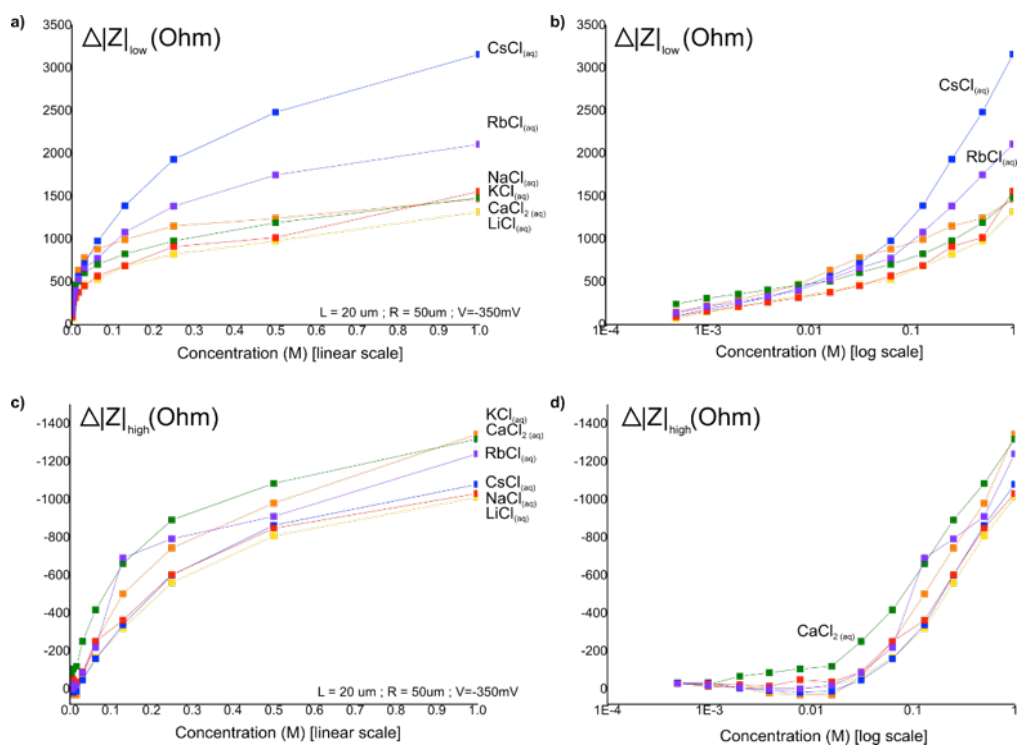

**Fig. S5**. Concentration dependency of the low-frequency impedance modulus variations (a and b) and high-frequency impedance modulus variations (c and d) of six different cations chloride aqueous electrolytes under $V_{DC} = -350$ mV. Abscissa in linear scale (a and c) and logarithmic scale (b and d) for the ion concentration.

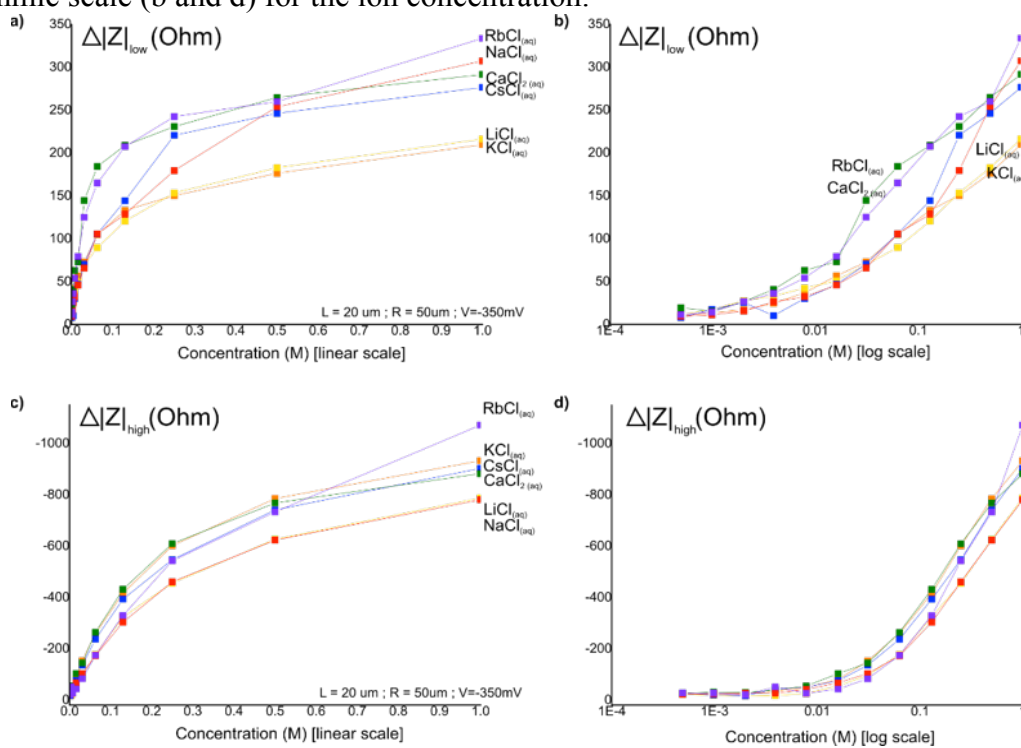

**Fig. S6**. Concentration dependency of the low-frequency impedance modulus variations (a and b) and high-frequency impedance modulus variations (c and d) of six different cations chloride aqueous electrolytes under $V_{DC} = -50$ mV. Abscissa in linear scale (a and c) and logarithmic scale (b and d) for the ion concentration.

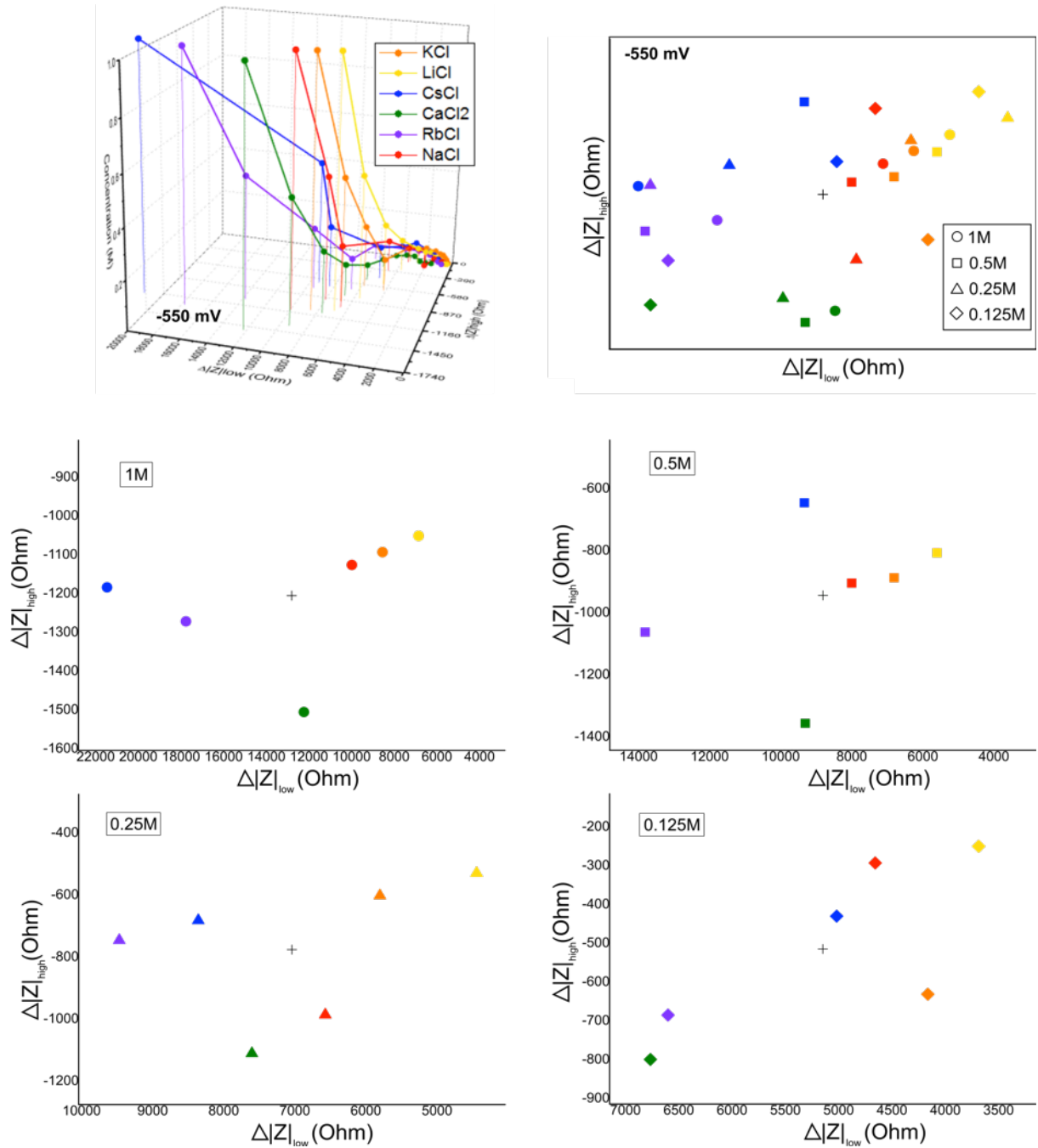

**Fig. S7**. Three dimensional plot of the low- and high-frequency impedance modulus variations with the ionic concentration under $V_{DC}$ = -550 mV. Stacking of 1 M, 0.5 M, 0.25 M and 0.125 M planes, centered on their respective average point (cross point). Graphs of each concentration plane at 1 M; 0.5 M; 0.25 M and 0.125 M, function of $\Delta|Z|_{low}$ and $\Delta|Z|_{high}$.

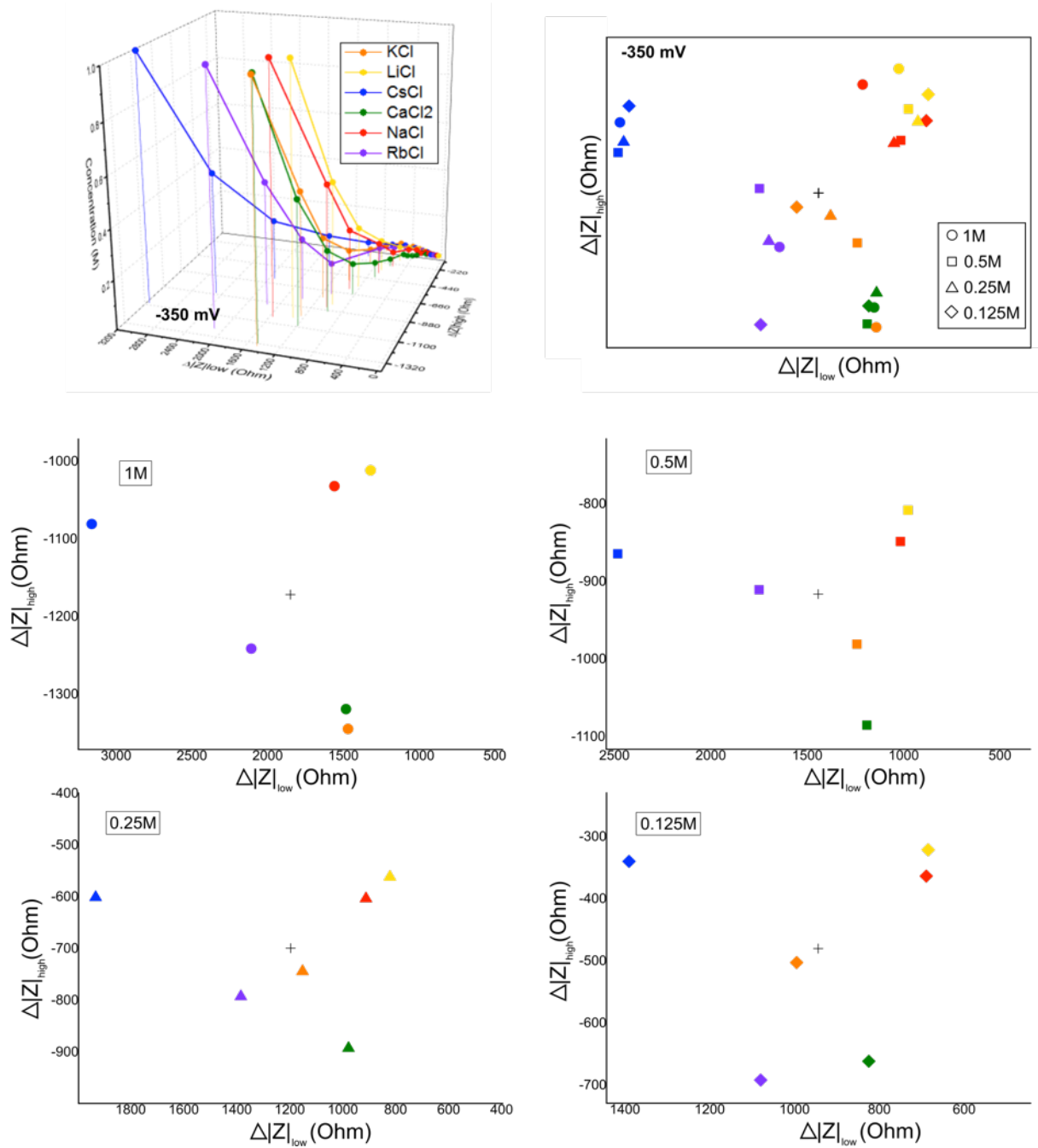

**Fig. S8**. Three dimensional plot of the low- and high-frequency impedance modulus variations with the ionic concentration under $V_{DC}$ = -350 mV. Stacking of 1 M, 0.5 M, 0.25 M and 0.125 M planes, centered on their respective average point (cross point). Graphs of each concentration plane at 1 M; 0.5 M; 0.25 M and 0.125 M, function of $\Delta|Z|_{low}$ and $\Delta|Z|_{high}$.

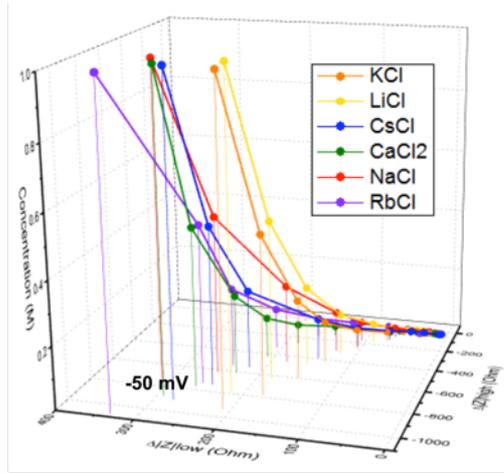
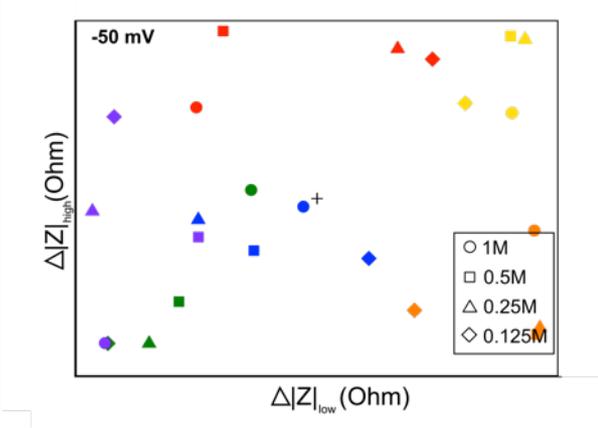
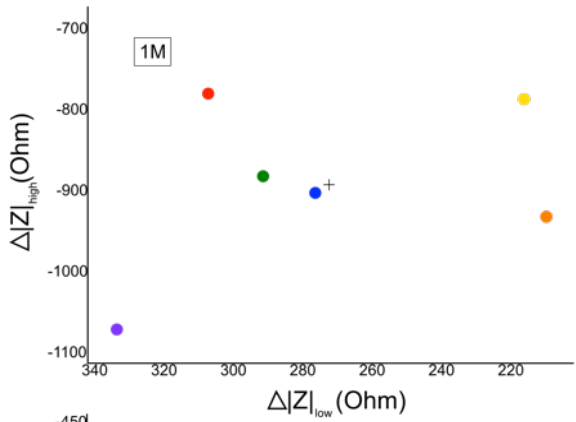
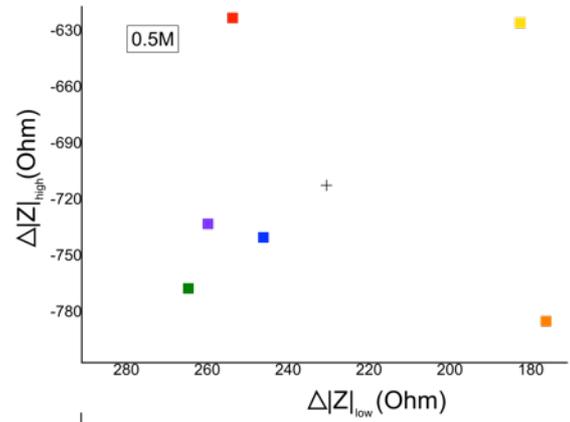
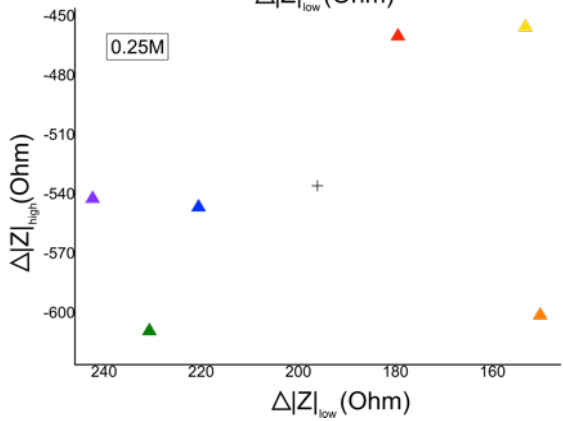
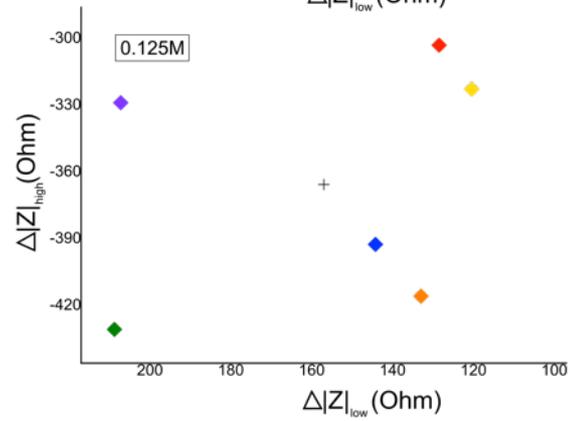

**Fig. S9**. Three dimensional plot of the low- and high-frequency impedance modulus variations with the ionic concentration under $V_{DC}$ = -50 mV. Stacking of 1 M, 0.5 M, 0.25 M and 0.125 M planes, centered on their respective average point (cross point). Graphs of each concentration plane at 1 M; 0.5 M; 0.25 M and 0.125 M, function of $\Delta|Z|_{low}$ and $\Delta|Z|_{high}$.